# Fixed exit DC-monochromator *of general position* for Side (or Top) Beam Line

N.G. Gavrilov[a], M.A. Sheromov[a], B.P. Tolochko[b], I.L. Zhogin[b,*]

[a] Budker Institute for nuclear physics - Lavrentjev 11 - 630090 Novosibirsk – Russia

[b] Institute for solid state chemistry - Kutateladze 18 - 630128 Novosibirsk - Russia

**Abstract**

We develop an idea of fixed exit double-crystal monochromator *of general position*, in which the lines of entering and exit beams (both being fixed) do not lie in one plane, i.e., are in general position.

Both cases of identical and not identical crystals are considered, and exact solutions are obtained describing valid positions and orientations of both crystals. Special attention is paid to the case of small lattice mismatch between the first and the second crystal (e.g. due to temperature difference) and small angle of "skewness" (that is to say, an angle between skew lines of entrance and exit beams is much less than unity).

We derive also some useful formulations describing beam profile change after two reflections (whose reflection planes are non-parallel). At last, possible designing approach for such a skew DC-mono is briefly touched upon.

*****Corresponding author:**   Zhogin I., Institute of Solid State Chemistry, Kutateladze st. 18, 630128, Novosibirsk, Russia

**Phone:**  +7(3832)394298     **E- mail:**  zhogin@inp.nsk.su

# 1. Introduction

Side beam lines of synchrotron radiation use as a rule single crystal monochromator providing monochromatic beam of a fixed energy (or beam energy can vary only in a very small range) [1]. The new functionality of energy tuning in a large range would be an important feature raising user value of a side beam line (and multiplying the use of an insertion device and storage ring as well). The possible way to attain this feature is a fixed exit double crystal monochromator of *general position* (or skew DC-mono where lines of entering and exit beams do not lie in one plane) considered in this communication. As we will show, this gives the opportunity that the first and the second crystal to be not identical.

Therefore, such a crystal pair as Diamond (111) and Ge(220), or Si(hkl) and Ge(hkl) could be used – without third optical element (mirror or multilayer) required in plane arrangement in order to compensate the lattice mismatch between crystals [2]. The desire to use not identical crystals is explainable by the fact that the first and the second crystals are working under very different thermal loads and should meet different requirements (especially when the second crystal is in use for sagittal focusing).

As we will see, in (slightly) skew monochromator with equal crystals, small changes of lattice spacing because of thermal expansion of the first crystal could be compensated through small corrections in crystals' positions and orientations. We believe that radiant cooling of (or heat abstraction from) the first crystal could be preferable in some conditions, especially having in view forthcoming synchrotron radiation sources of forth generation.

Such a not-planar approach to DC-mono setup opens up evident possibility to have a set of skew monochromators shedding rays of different energy on the same sample.

# 2. Skew DC-monochromator (with fixed exit)

Let us consider two skew lines in Figure 1, $x$ and $x'$, which indicate incoming and outgoing light beams and serve as guide lines (or rail guides) for two ideal mirrors placed at points $A$ and $A'$. It is possible to orientate the first mirror in such a way that to throw reflected light spot to the point $A'$; in turn, the second mirror can be oriented to redirect double-reflected light beam along outgoing line $x'$.

If mirrors are not ideal, but instead some ordinary crystals serve to reflect x-rays, then their angles of reflection (Bragg angles) should be coordinated to match their pass bands. This requirement leads to one equation for $x$ and $x'$ having a form $F(x, x') = 0$. So the crystal positions should be coordinated according to this equation.

Let the minimal length between skew lines $x$ and $x'$ (i.e., between points $O$ and $O'$, see Figure 1) is taken as a unit of length: $h = 1$. Then we obtain the following coordinates for the points $A$ and $A'$:

$$A = (x, 0, 0), \qquad A' = (-x' \cos\alpha,\ x' \sin\alpha,\ 1); \tag{1}$$

here $x$ (and $x'$) is a length measured from the origin $O$ (and $O'$), that is a length of the line segment $A - O$; $\alpha$ is a skew angle formed by projection of skew lines along the segment $O - O'$.

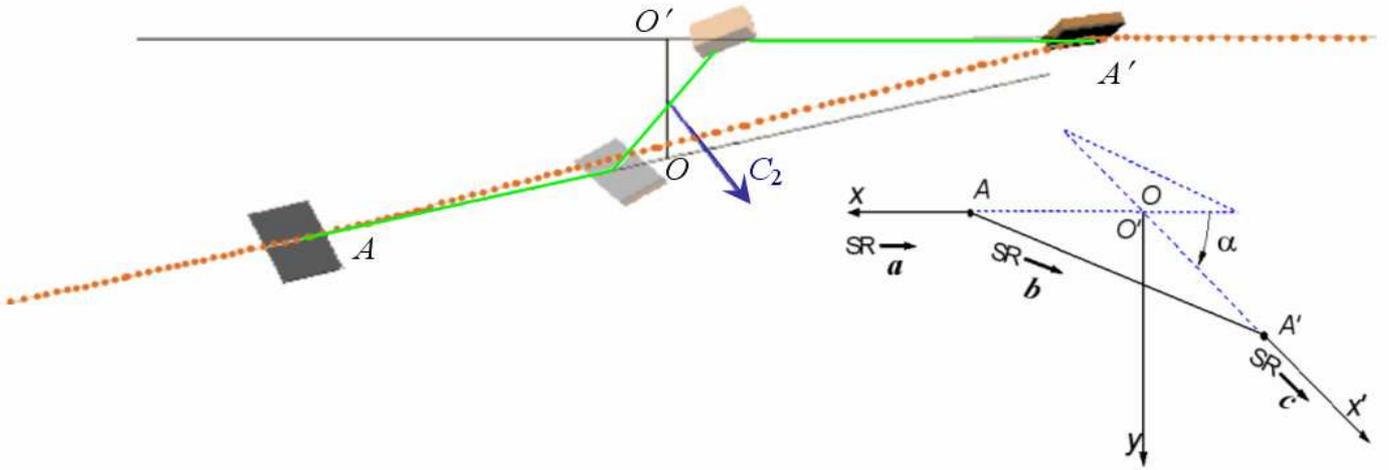

Figure 1. Fixed exit DC-mono of general position (skew mono) – illustrating outline and top view.

Using (1) we find the unit vectors along parts of beam trajectory, see Figure 1:

$$\vec{a} = (-1,\ 0,\ 0),\quad \vec{b} = \frac{(-x - x'\cos\alpha,\ x'\sin\alpha,\ 1)}{\sqrt{x^2 + x'^2 + 2xx'\cos\alpha + 1}},\quad \vec{c} = (-\cos\alpha,\ \sin\alpha,\ 0). \tag{2}$$

The equations for Bragg angle $\theta$ and roll angle $\varphi$ (conceived as the angle between reflection plane,

$$\{\vec{a},\ \vec{b}\},\quad (0,\ x'\sin\alpha,\ 1) \in \{\vec{a},\ \vec{b}\},$$

and $xz$–plane) of the first crystal are now obtained easily using Eq. (2):

$$2\sin^2\theta = 1 - (\vec{a}\cdot\vec{b}) = 1 - \frac{x + x'\cos\alpha}{\sqrt{x^2 + x'^2 + 2xx'\cos\alpha + 1}},\qquad \tan\varphi = x'\sin\alpha. \tag{3}$$

Similar equations are valid for primed angles of the second crystal (it is necessary just to replace primed and unprimed values excepting $\alpha$):

$$2\sin^2\theta' = 1 - (\vec{c} \cdot \vec{b}) = 1 - \frac{x' + x\cos\alpha}{\sqrt{x^2 + x'^2 + 2xx'\cos\alpha + 1}}, \quad \tan\varphi' = x\sin\alpha. \quad (3')$$

The pass bands of two crystals will be matched if

$$\lambda = \lambda', \text{ where } \lambda = 2d\sin\theta, \quad \lambda' = 2d'\sin\theta',$$

or

$$\sin^2\theta = q^2 \sin^2\theta', \text{ where } q = d'/d. \quad (4)$$

Here, in Eq. (4), $d$ ($d'$) is a lattice spacing of the first (second) crystal.

This match condition fixes $x'$ as a function of $x$ (or vice versa). However, there are values of parameters $\alpha$ and $q$, that solution of Eq. (4) is absent: $\alpha = 0$, $q \neq 1$. Substitution of these values to Eq. (4) leads to the equation which obviously has no solution (well, in real numbers):

$$\sqrt{(x+x')^2 + 1} = x + x'.$$

On the other hand, the special case $\alpha = 0$, $q = 1$ admits any combination $x, x'$ as solution of match condition, Eq. (4); this is a feature concerned with additional symmetry of whole system – translations along parallel lines $x$ and $x'$ (which lie now in one plane). It is due to this extra-symmetry that one crystal of the pair can be fixed (that is to have fixed position) as it takes place in usual (i.e. plane) fixed-exit double-crystal monochromator (in nondispersive setting). It is obvious, however, that this special case with its extra-symmetry is not stable with respect to small deviation $\delta d = d - d'$ (due to temperature difference between crystals and their thermal expansion).

## 2.1. *Identical Crystals*

In the case of equal crystals, when $d = d'$, we have easy solution of Eq. (4):

$$x = x', \quad \theta' = \theta, \quad \varphi = \varphi',$$

and hence [see Eqs. (3), (3′)]

$$\sin\theta = \sqrt{\frac{1}{2} - \frac{x(1+\cos\alpha)}{2\sqrt{2x^2(1+\cos\alpha)+1}}}, \quad \varphi = \arctan(x\sin\alpha). \qquad (5)$$

This solution keeps the symmetry $C_2$ with symmetry axis being the bisector (placed at the middle of $O$–$O'$) of adjacent angle, $\pi - \alpha$; see Figure 1.

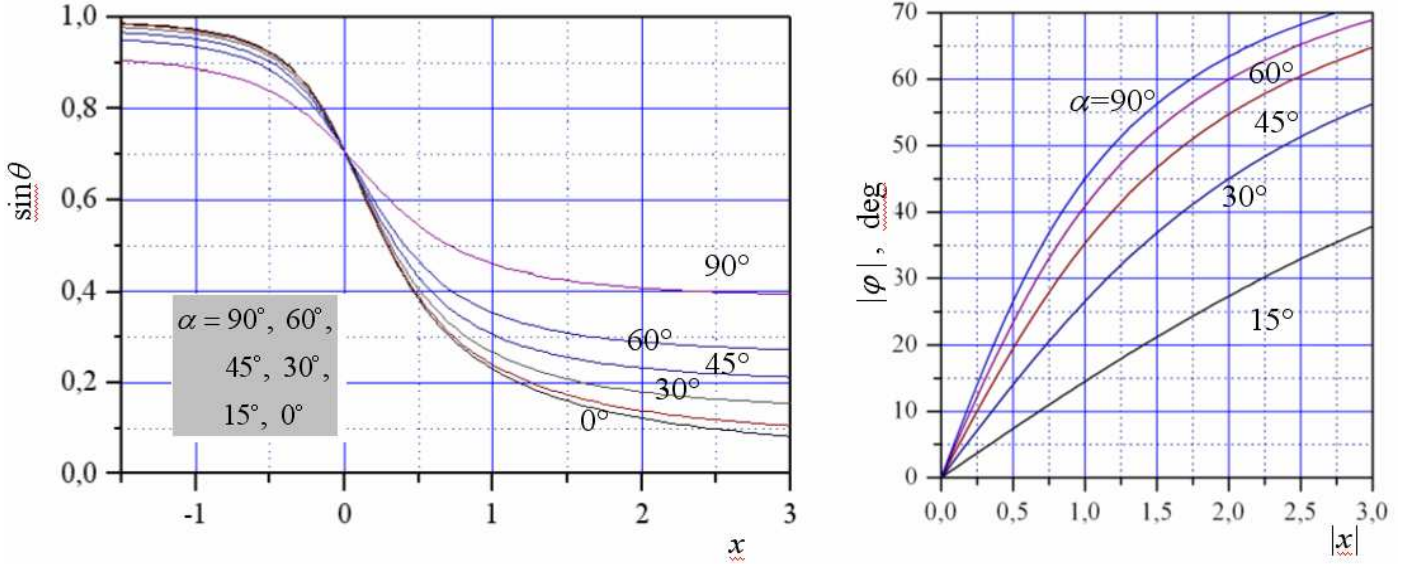

Figure 2. (a) Plots $\sin\theta(x;\alpha)$ and (b) $\varphi(x;\alpha)$ calculated from Eq. (5).

Figures 2a and 2b show $\sin\theta(x)$ and $\varphi(x)$, respectively; Eq. (5) is in use with different values of $\alpha$. For small $\alpha, x$ ($\alpha x, \alpha < 1$ with $\alpha$ in radians, of course) we have approximate equations:

$$\theta(x;\alpha) = \frac{1}{2}\arctan\frac{1}{2x} + \frac{\alpha^2 x}{4}\frac{1+2x^2}{1+4x^2}\left(1 - \frac{\alpha^2}{12} - \frac{\alpha^2 x^2}{2}\frac{2+7x^2+4x^4}{1+6x^2+8x^4}\right) + O(x\alpha^6, x^5\alpha^6);$$

$$\varphi(x;\alpha) = x\alpha\left(1 - \alpha^2\frac{1+2x^2}{6} + \alpha^4\frac{1+20x^2+24x^4}{120}\right) + O(x^7\alpha^7, x\alpha^7).$$

Considering limits $x \to \pm\infty$ in Eq. (5), one can find the possible range of Bragg angle depending on the skew angle $\alpha$:

$$\theta_- = \theta_{\min} = \alpha/4, \quad \theta_+ = \theta_{\max} = \pi/2 - \alpha/4, \quad \Delta\theta = (\pi - \alpha)/2. \qquad (6)$$

So we see that an increase in skew angle $\alpha$ leads to decrease in the range of available Bragg angle. In general case ($q \neq 1$), this range $\Delta\theta$ will also depend on $q$; above-mentioned absence of solution of Eq. (4) in the case $\alpha = 0, q \neq 1$ means that this available range is empty.

## 2.2. *Not identical Crystals*

In general case of not identical crystals, $\alpha \neq 0$, $q \neq 1$, Eq. (4) yields (using Eqs. (3) and (3′) with cancellation of factor $1-q^2$)

$$\sqrt{x^2 + x'^2 + 2xx'\cos\alpha + 1} = x + x' + \kappa(qx - x'/q), \qquad (7)$$

we introduce here parameter $\kappa = (1-\cos\alpha)/(1/q - q)$. This equation (7) still has symmetry – namely with respect to the replacement

$$x \leftrightarrow x', \quad q \to 1/q \quad (\kappa \to -\kappa).$$

Because of this symmetry we will assume further that $q < 1$ and hence $\kappa > 0$.

Elimination of the radical in Eq. (7) leads to the following quadratic:

$$x^2 \kappa q(\kappa q + 2) - 2\kappa^2 xx' - x'^2(2 - \kappa/q)\kappa/q - 1 = 0. \qquad (8)$$

Using $x'$ as independent (or control) variable, one can find the exact solution:

$$x = x(x'; q, \alpha) = \frac{\kappa x' + \sqrt{q(q + 2/\kappa) + 4x'^2 \cos^2(\alpha/2)}}{q(2 + \kappa q)}, \quad \text{where} \quad \kappa = q\frac{1-\cos\alpha}{1-q^2}; \qquad (9)$$

here only the positive branch is correct – superfluous brunch arises in Eq. (8) because of previous radical elimination in Eq. (7).

Considering the limits $x' \to \pm\infty$ in Eq. (9), we get asymptotic coefficients, $\gamma_\pm$, and the limits, $\theta_\pm$, of available Bragg angle range (instead of Eq. (6) for identical crystals):

$$\gamma_\pm = \lim_{x' \to \pm\infty} \frac{x}{x'} = \frac{\kappa \pm 2\cos(\alpha/2)}{q(2 + \kappa q)}, \quad \tan 2\theta_\mp = \pm\frac{\sin\alpha}{\cos\alpha + \gamma_\pm}. \qquad (10)$$

Note that large value of parameter $\kappa$ (with $q \approx 1$) leads to $\gamma_\pm \approx 1$, and the last equation in (10) yields the range of Bragg angle close to that in Eq. (6) which is valid for the case of identical crystals (with $x' = x$ and $\gamma_\pm = 1$, naturally).

We will consider now two particular cases of crystal choice and parameter set.

(1) Crystal pair: Diamond(111) and Ge(220); $d = 2.059$, $d' = 2.00$; $q = 0.971$.

  Skew angle: $\alpha = 0.3$ (17.2°), $\kappa = 0.77$.

(2) Identical crystals having different temperature: $\delta q = 1 - q \approx 5 \cdot 10^{-4}$ ($\Delta T \approx 200°$ for Silicon).

  Skew angle: $\alpha = 0.1$ (5.7°), $\kappa = 5$.

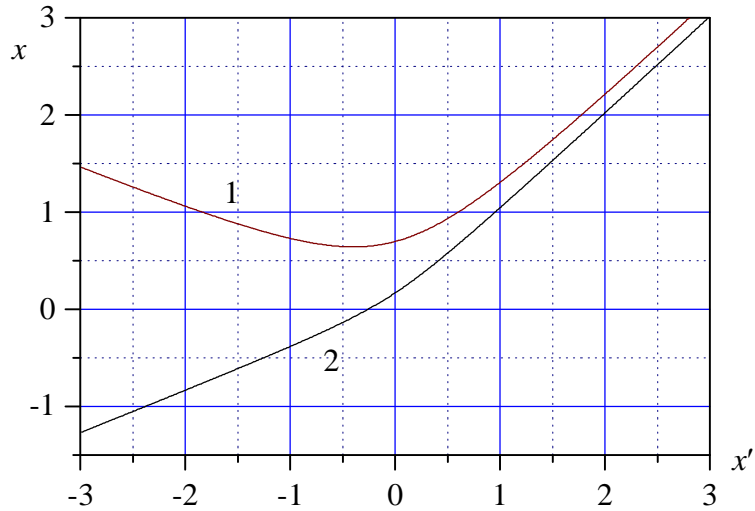

Figure 3. Plots $x(x')$ for the two choices of parameters (see in text).

Plotted functions $x(x')$ and $\sin\theta(x')$ for these choices of parameter set are shown in Figure 3 and Figure 4, respectively.

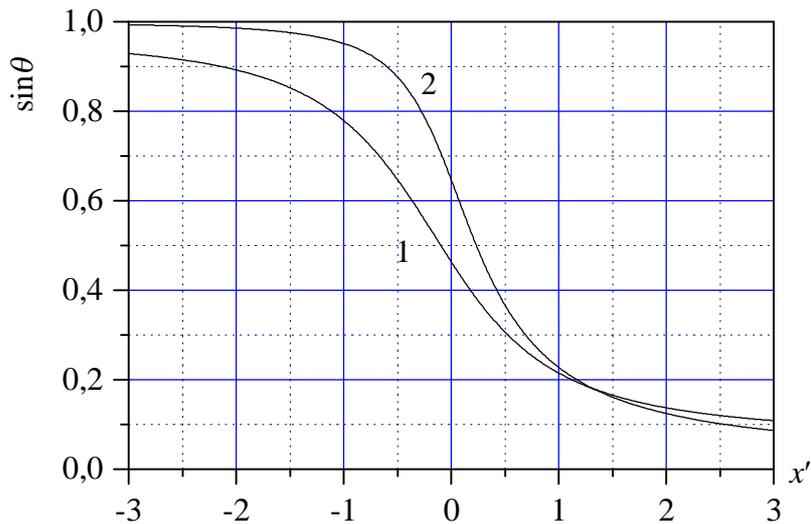

Figure 4. Plots $\sin\theta(x')$ for the two choices of parameters.

## 3. Beam Profile issue

Let us assume that the initial white beam has fan-like form, and its width is much greater than its height. As a result of two reflections in skew monochromator, the beam profile line (and beam polarization as well, but it is some more complicated issue) will be turned on a small angle, $\psi_l$.

Calculation of this angle is the aim of this section. Let us consider sphere of unit radius – the ends of vectors $\vec{a}, \vec{b}, \vec{c}$ are designated (their starting points coincide with the centre of sphere, of course), see Fig. 5a. The arcs of spherical triangle $\triangle_S(abc)$ have the following lengths:

$$\widehat{ab} = 2\theta, \quad \widehat{bc} = 2\theta', \quad \widehat{ac} = \alpha \ .$$

We need to find the angles of this triangle, $\varphi_a, \varphi_b, \varphi_c$, where $\varphi_a$ is the angle at the triangle vertex $a$. Note that the case of small angle $\varphi_b$ (and large $\theta, \theta'$) is close to nondispersive (+,-)-setting, while large value of this angle ($\varphi_b \to \pi$) corresponds to dispersive (+,+)-setting.

Consider unit vectors $\vec{\tau}_{ab}, \vec{\tau}_{ac}$, tangent at vertex $a$ to arcs $\widehat{ab}$ and $\widehat{ac}$, respectively, see Fig. 5b. It is easy to find (using $(\vec{\tau}_{ab} \cdot \vec{a}) = 0$ and so on) that

$$\vec{\tau}_{ab} = \frac{\vec{b} - \vec{a}(\vec{a} \cdot \vec{b})}{\sqrt{1 - (\vec{a} \cdot \vec{b})^2}}, \quad \vec{\tau}_{ac} = \frac{\vec{c} - \vec{a}(\vec{a} \cdot \vec{c})}{\sqrt{1 - (\vec{a} \cdot \vec{c})^2}} \ ;$$

hence we are now about to obtain $\varphi_a$ (and other angles – in a similar way):

$$\cos \varphi_a = (\vec{\tau}_{ab} \cdot \vec{\tau}_{ac}) = \frac{(\vec{b} \cdot \vec{c}) - (\vec{a} \cdot \vec{b})(\vec{a} \cdot \vec{c})}{\sqrt{1 - (\vec{a} \cdot \vec{b})^2}\sqrt{1 - (\vec{a} \cdot \vec{c})^2}} = \frac{\cos 2\theta' - \cos 2\theta \ \cos \alpha}{\sin \alpha \ \sin 2\theta} \ . \tag{11}$$

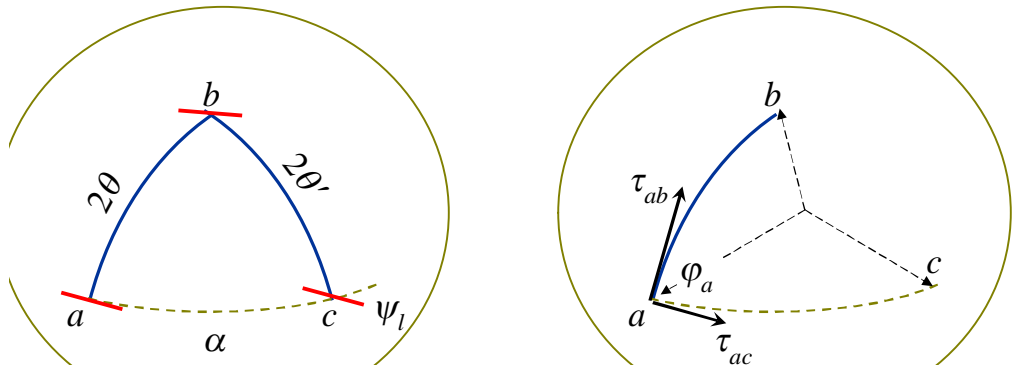

Figure 5. (a) Spherical triangle on unit sphere; (b) tangent vectors at vertex $a$.

The strokes at triangle vertexes in Fig. 5a indicate beam profile line, and this property will be considered in the following subsection.

It is easy to realise that the angle between the beam stroke and reflection arc (e.g. arc $\widehat{ab}$ for the first reflection) does not change during reflection; in other words, the stroke is parallel translated along arc of reflection which is geodesic line on the unit sphere. The result of two translations along arcs $\widehat{ab}$ and $\widehat{bc}$ will some differ from the result of single parallel translation along the arc $\widehat{ac}$, and this difference is equal to the target angle $\psi_l$. This angle has simple relation to the angles of spherical triangle, Eq. (11), or to its area ($S_{abc}$):

$$\psi_l = \varphi_a + \varphi_b + \varphi_c - \pi = S_{abc}. \tag{12}$$

Very simple (and very crude) estimate for this value is as follows:

$$\psi_l \sim \alpha\,\theta \quad (\theta \sim \theta'). \tag{13}$$

In the case of identical crystals (isosceles spherical triangle), one can use Eqs. (11), (12) to obtain the next explicit equation:

$$\sin(\psi_l/2) = \tan(\alpha/2)\frac{\sqrt{\sin^2 2\theta - \sin^2(\alpha/2)}}{1+\cos 2\theta}; \tag{14}$$

one can see that the Eq. (14) leads to the estimation (13) if $\alpha < \theta < 1$.

## 4. Possible Design Approach

Figure 6 shows the kinematics of ultra-high vacuum DC-monochromator (with *practically* fixed exit) which is under construction now in Siberian SR centre [3]. (Here "practically" means that relative deviation $\Delta h/h$ is about $3\cdot 10^{-4}$ for Bragg-angle range $5° < \theta < 18.5°$. Less range gives less deviation.) The source of movement in this kinematics is linear (not vacuum) positioner, and all the most crucial surfaces of parts of construction are very simple, plane or cylinder surface.

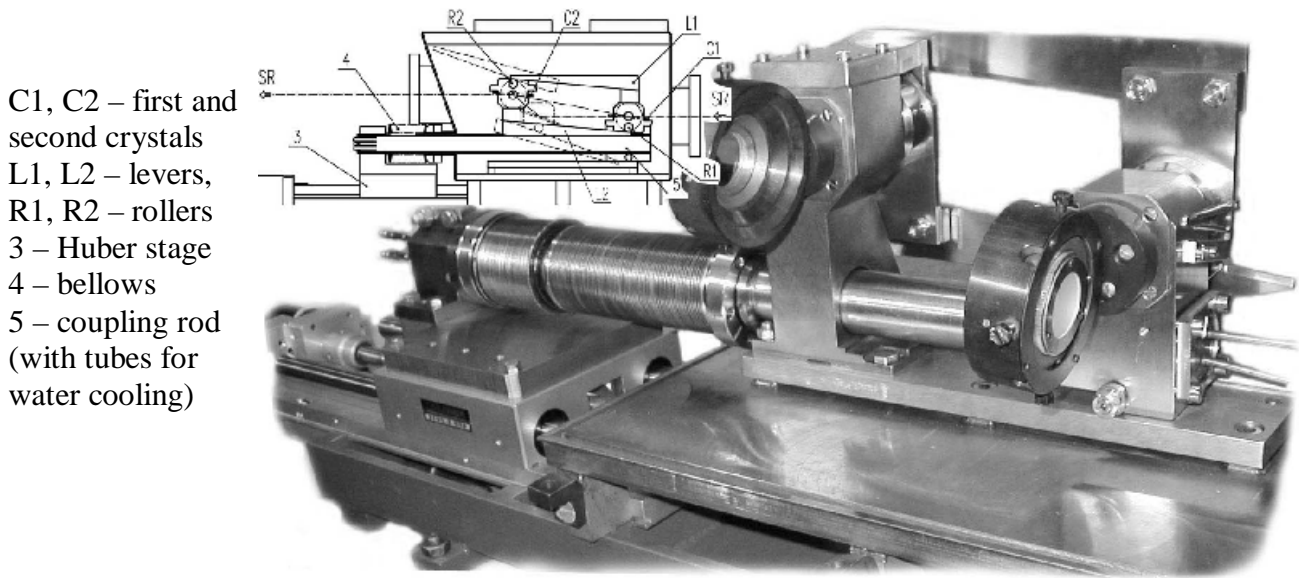

C1, C2 – first and second crystals
L1, L2 – levers,
R1, R2 – rollers
3 – Huber stage
4 – bellows
5 – coupling rod (with tubes for water cooling)

Figure 6. Kinematics of the plane DC-mono (mirrors instead of crystals; $\theta \to \pi/2 - \theta$; $h$=40) [3].

We believe that similar approach with two linear positioners could be applied to design of a skew DC-monochromator. There is a dilemma which angle is the first – $\theta$-angle or $\varphi$-angle, see Eqs. (3), (3′) and Fig. 1. One choice is outlined in the Figure 7:

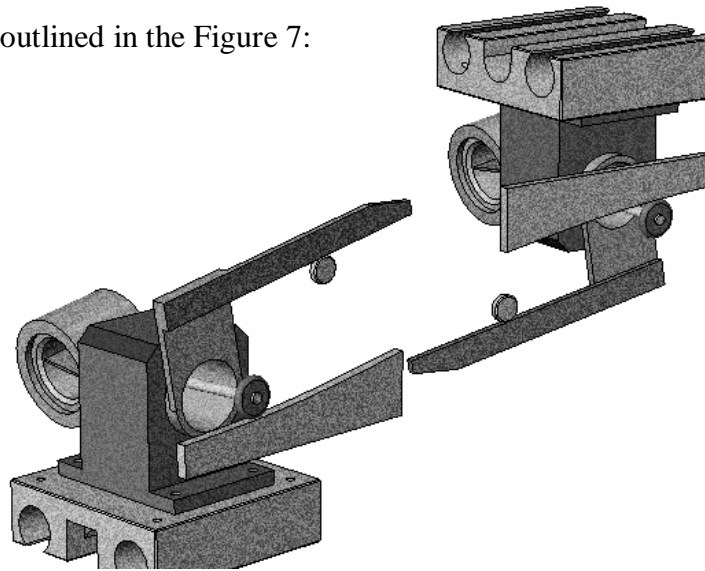

Figure 7. Sketch of possible kinematics of a skew DC-monochromator.